# Exploring Vibrotactile Intensity Perception with Multiple Waveform Parameters

Takumi Kuhara[1], Hikari Yukawa[1], and Yoshihiro Tanaka[1,2]

[1]*Dept. of Electrical and Mechanical Engineering, Graduate School of Engineering,*
*Nagoya Institute of Technology, Japan*

[2] *Inamori Research Institute for Science, Japan*

(Email: t.kuhara.538@nitech.jp)

**Abstract ---** It is known that by longing the duration of a vibrotactile stimuli or applying a damping or an increasing factor to the waveform the perceived intensity is affected in different ways. This paper presents a vibrotactile presentation system assembled with a software waveform generator that enables comparison in the perceived intensity for different waveforms made by multiple parameters. The adjustable parameters are frequency, amplitude, and wave type for the basic part of the stimuli and in addition, it is possible to apply an exponential decay or increasing factor to the waveform by specificizing the duration. By using the presented system, an easy comparison of the influence on the perception of intensity by different parameters of the waveform is possible. We conducted a preliminary experiment on a variety of waveshapes with and without damping by using this system.

## 1 INTRODUCTION

Due to the development of technology, various principles of tactile sensations have been elucidated and have various usages. For virtual reality, providing tactile information that strengthens the recognition of the visual stimuli improves the immersive feeling [1] and reduces the awkwardness that comes from the difference from reality, leading to more natural interaction between a virtual object and the person [2].

When providing tactile information in an application, vibrotactile information is often used because of its easiness to reproduce the spatiotemporal difference and to present various stimuli with different characteristics. For vibrotactile stimuli, there are a variety of parameters that affect the feeling when presented. For example, the shape of the waveform, frequency, and amplitude are all important factors that form the basic feeling provided as stimuli. It is reported that the frequency and amplitude of the vibrotactile feedback for 2D navigation affected the performance and comfort [3]. The amplitude and frequency of vibrotactile stimuli are also reported to affect the intuitiveness and effectiveness of the feedback provided [4]. We aim to investigate the influence of multiple parameters on intensity perception for vibrotactile stimulation. In this paper, we introduce an assembled vibrotactile presentation system that consists of a software waveform generator that enables people to generate different waveforms using multiple parameters for an easy comparison of perceived intensity and a preliminary experiment conducted by using this system.

## 2 ASSEMBLED WAVEFORM GENERATOR

We developed a software waveform generator using Python 3.10 as shown in Fig. 1. The top half shows the waveform that will be calculated using the parameters shown in the bottom half of the GUI which is all going to be calculated with a sample frequency of 44.1 kHz.

For the shape of the waveform, there are four choices (the sinusoidal wave, square wave, triangle wave, and sawtooth wave) prepared and users can choose one from the pulldown for the shape of the waveform. The frequency and duration are adjustable by inputting a desired value for each parameter, elsewhere they are set with the default values of 200 Hz and 0.3 s. The amplitude is also changeable by a slider or input with a minimum of 0.0 and a maximum of 1.0 which is considered 0 and 100 for the keyboard input. Once pressing "generate" on the middle left, the calculated waveform is played on the chosen output index and type continuously or having a 0.1 s interval between each playback if desired. Additionally, we made it possible to apply a decay or increasing factor to the waveform. The exponent of the damping and increasing formula is adjustable according to the duration set, allowing changes to the damping or increasing shape of the waveform.

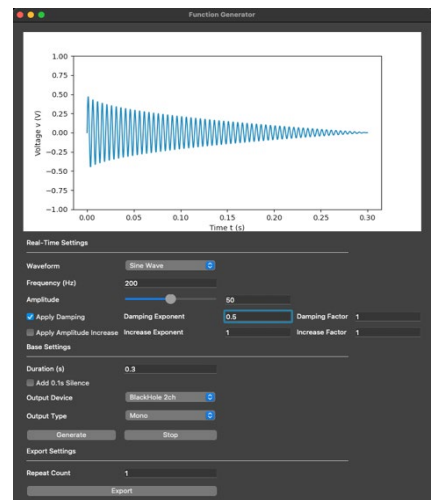

**Fig 1. GUI of the waveform generator**

### 3 COMPARISON OF MULTIPLE WAVEFORMS

To see the difference in intensity perception, we made a comparison between different waveforms with different shapes and with a difference in the appliance of the damping factor, which were all made with the same duration and frequency.

#### 3.1 Stimuli

Five samples were exported using the application: a sinusoidal wave, a damping sinusoidal wave, a square wave, a damping square wave, and a damping saw tooth wave. Each sample was made with a sampling frequency of 44.1 kHz, a frequency of 200 Hz, a duration of 0.5 s, and repeated 5 times with a 0.1 s silent sound in between. The sinusoidal wave and the damping sinusoidal wave each exhibited a maximum acceleration of 5.6 m/s². In contrast, the damping sawtooth wave reached a maximum acceleration of 15.1 m/s², while both the square wave and the damping square wave attained maximum accelerations of 19.9 m/s².

#### 3.2 Experiment Method

For the experiment, we used an amplifier (Foster, AP05 mk2) and a vibrotactile actuator (Foster, 646752) to present the stimulation. Six participants (5 males, and 1 female) were instructed to rank each sample in the order of perceived intensity. The participants were presented each sample once in a random order and after ranking all samples, they were presented the samples once again in the order of their answered rank to check whether their evaluation was appropriate and re-ordered if they wanted.

#### 3.3 Results and Discussion

The results are shown in Fig. 2 as a boxplot. The horizontal axis shows the rank answered for each sample. The answered ranks of the damping sawtooth wave were close to the results for the damping sinusoidal wave. The results for the damping square wave were close to those for the sinusoidal wave and the square wave was answered the strongest. Regarding the relationship between perceived intensity and acceleration, the sensory rankings for constant waveforms, specifically sinusoidal and square waves, corresponded directly with their respective acceleration magnitudes. In contrast, for damping waveforms, while the square wave continued to be perceived as the most intense, the sawtooth and sinusoidal damping waves were perceived with nearly equivalent intensities. Additionally, all damping waveforms were consistently perceived as less intense compared to their constant waveform counterparts, regardless of the measured acceleration values.

This result insists that damping waveforms are perceived as having lower intensity. The results of the sinusoidal wave and the square wave were consistent with the results reported by Verrillo [5], which demonstrated that the square wave was perceived as the strongest, the sawtooth wave next, and the sinusoidal wave last. For the decaying waveforms, the reports from Gescheider et al. [6] discussed that the perception of the damping waveform should not affect the above-mentioned order. The results of our experiment did not support their conclusion; however, this may be caused by the difference in the damping shape and the limitations of this experiment such as the number of participants and the characteristics of the vibrator that we used to present the stimulation. To provide additional insight into the relationship between waveform characteristics and perceived intensity, another factor to consider is the time-averaged acceleration of each waveform in relation to perceived intensity. For damping waveforms, the acceleration varies throughout the duration of the waveform, which may have influenced their perceived intensity. Also, the previous study discussed the possibility of adaptation affecting the results for decaying waveforms. For the experiment, we used waveforms with a duration of 0.5 s which may be short against the adaptation.

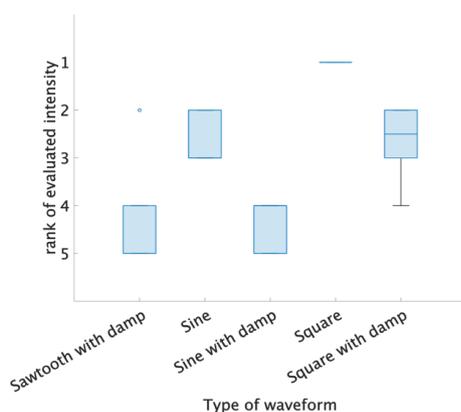

**Fig 2. Results for the Experiment**

## 4 Conclusion

We presented a system that consists of an assembled software waveform generator that enables easy comparisons of perceived intensity between different parameters for vibrotactile stimuli. For future research, we will investigate how much each parameter affects tactile sensations.

### Acknowledgment

This work was supported by JST SPRING, Grant Number JPMJSP2112, JSPS Grant in Aid for Early Career Scientists Grant Number 24K20816, and Inamori Research Institute for Science.